# A Lightweight Ethereum Voting Prototype for Hospital Ethics Committees with Receipt-Based Inclusion Verification

Edwin Clatus
The Pennsylvania State University
University Park, PA, USA
edwinc3173@gmail.com

Madhusudhan Singh
The Pennsylvania State University
University Park, PA, USA
mps6990@psu.edu

***Abstract*—**This paper presents a Solidity, Hardhat, React, MetaMask, and ethers.js prototype for hospital ethics-committee voting. Role controls, case-state checks, duplicate-vote controls, and a receipt hash support public audit and transaction-inclusion verification. Because vote events expose wallet addresses and vote values, the design provides pseudonymous auditability, not anonymous or secret-ballot voting; the receipt is neither receipt-free nor coercion-resistant. Evaluation reports 22 passing functional tests and local Hardhat gas use, including 284,137 gas per vote. A 12-participant simulation used assumed probabilities and is not human-subject evidence. Residual risks include multiple wallets, administrator or frontend compromise, credential reassignment, front-running, denial of service, and untested adversarial paths. Confidential deployment requires governed enrollment, encrypted ballots, independent audit, adversarial testing, reproducible benchmarks, and a real user study.

## I. INTRODUCTION

Clinical ethics committees require controlled procedures [1]. Paper ballots hinder independent audit; centralized systems require trust in the operator and logs; public blockchains provide durable records but conflict with ballot secrecy.

We ask whether an Ethereum contract can enforce eligibility and case state while allowing inclusion verification for a small committee. It cannot provide anonymity: addresses and votes are public, and institutions may link wallets to members. The claim is therefore pseudonymity, not confidentiality from observers or administrators.

The scope is a low-frequency prototype. Patient-identifying data must remain off-chain; deployment also requires institutional review, wallet recovery, and stronger ballot protection.

The contributions are:

- A full-stack, role-controlled Ethereum voting and case-lifecycle prototype.
- An independently reproducible inclusion-receipt procedure with an explicit public-data boundary.
- A threat model spanning observers, voters, frontends, administrators, multiple wallets, front-running, and denial of service.
- Evidence-bounded tests, gas, costs, synthetic inputs, and baselines.

## II. RELATED WORK

The blockchain-based voting systems have been widely established and used as mechanisms for improving transparency and auditability in digital elections. For example, Yang et al. [1] proposed a publicly verifiable blockchain voting protocol without trusted tallying authorities, demonstrating strong transparency guarantees but with higher implementation complexity, rooted in similarity to our concept. Additionally, Jafar et al. [2] surveyed blockchain e-voting systems and identified persistent challenges including privacy, scalability, and operational feasibility. Sharp et al. [3] further emphasized that many blockchain voting architectures remain impractical for institutional deployment due to cryptographic overhead and governance complexity, all issues our prototype has accounted for.

More recent systems have explored publicly verifiable voting without trusted tallying authorities is studied in [2]; surveys [3], [4] identify persistent privacy, scalability, key-management, usability, and governance limits that ledger immutability alone does not solve.

Stronger designs combine ring signatures and threshold sharing [5], blind signatures, zero-knowledge proofs, and threshold encryption [6], private boardroom smart contracts [7], or coercion-resistant receipts [8]. They improve privacy or coercion resistance at higher cryptographic and implementation cost.

Helios provides a centrally operated open-audit comparison [9], whereas an RBAC database is simpler but operator-dependent. This prototype uses public state and a Keccak-256 receipt without ballot encryption or zero-knowledge proofs, exposing wallet-vote linkage. Section V-E compares properties and marks the absence of a controlled same-platform benchmark.

## III. PROTOCOL AND SYSTEM DESIGN

### *A. Requirements and workflow*

The five phases are enrollment, case creation, vote submission, receipt verification, and resolution. React prepares the call, MetaMask signs it, ethers.js submits it, and the contract records state and an event that an independent verifier can retrieve by transaction hash.

Target properties are enrolled-wallet eligibility, open-case voting, one accepted ballot per wallet and case, and transaction-inclusion verification. The public representation does not provide ballot secrecy.

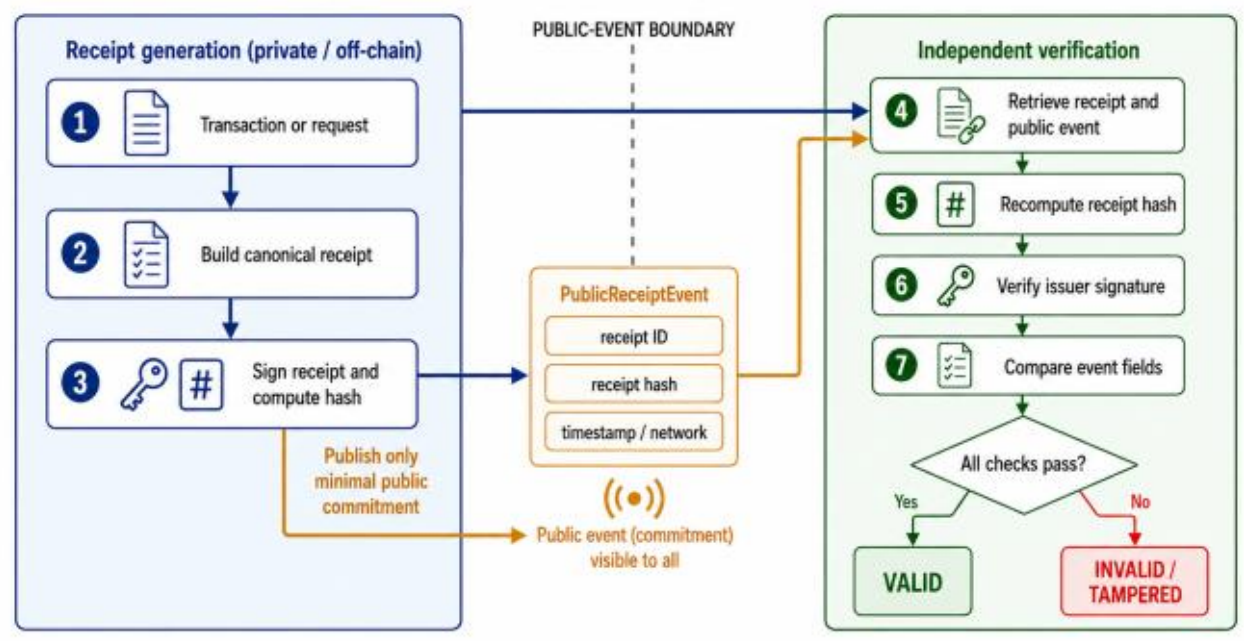

Fig. 1. Revised receipt-generation and independent-verification workflow. The public-event boundary is shown explicitly.

## *B. Receipt construction and independent verification*

The receipt is a deterministic hash of the accepted transaction fields. To avoid encoding ambiguity, the revised specification uses Solidity ABI encoding rather than an unspecified concatenation [Fig. 2]:

*receiptHash = keccak256(abi.encode(vote, caseId, voterAddress, blockTimestamp, nullifierHash)).*

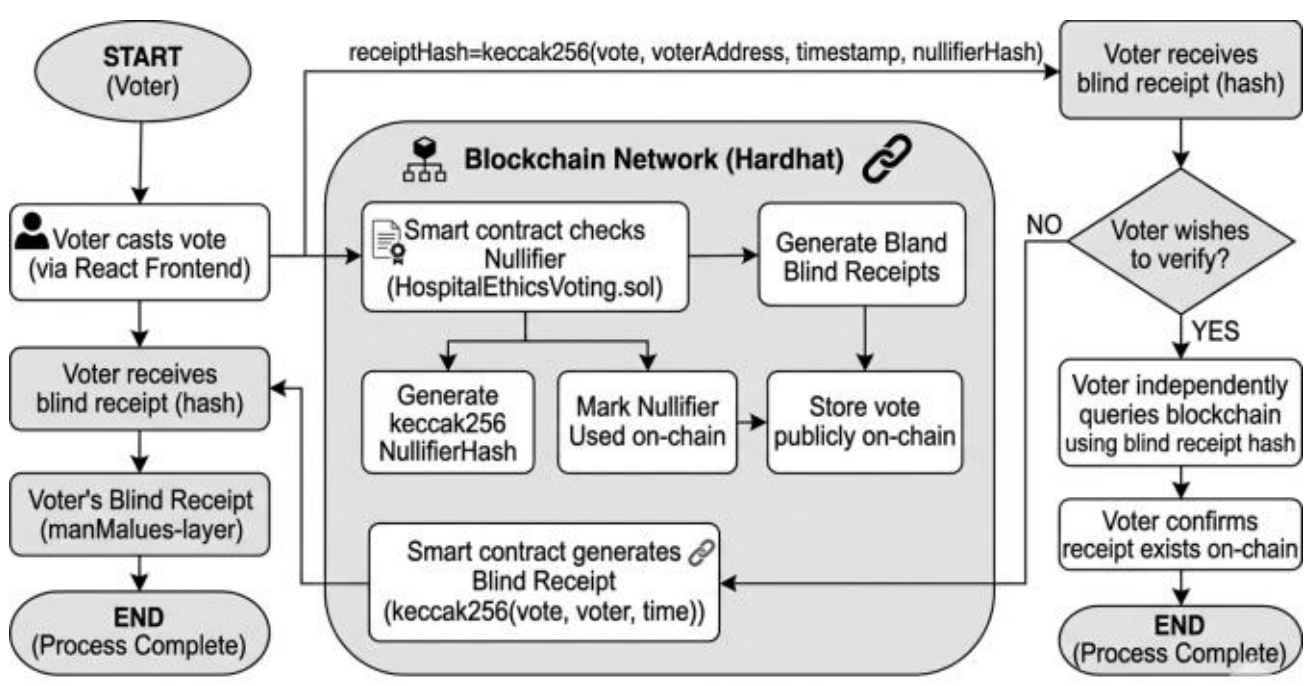

Fig. 2. Blind Receipt mechanism.

Independent recomputation requires every Table I input and the exact encoding from canonical transaction, event, block, or state data. Checking only whether a stored hash exists is an inclusion lookup, not recomputation.

TABLE I. RECEIPT PREIMAGE AND VERIFICATION SOURCE

| Field | Verifier source | Privacy implication |
|---|---|---|
| vote | Local record + event/state | Ballot observable |
| caseId | Call data or event | Public case linkage |
| voterAddress | Sender or event | Pseudonymous identity |
| blockTimestamp | Canonical block/event | Public metadata |
| nullifierHash | Event or state | Linkable audit tag |
| receiptHash | Local + event/state | Inclusion token only |

The voter verifies a receipt as follows:

1. Retain the transaction and displayed receipt hashes.
2. Query an independent provider for the canonical receipt, block, and vote event.
3. Extract and ABI-encode the five fields; recompute Keccak-256.
4. Compare local, recomputed, and on-chain values; any mismatch fails verification.

Because the event already exposes address and vote, this process adds no secrecy. The receipt neither proves eligibility independently nor provides receipt-freeness or coercion resistance [8], [10]; it verifies inclusion only.

## *C. Nullifier and duplicate-vote controls*

The nullifier is keccak256(abi.encode(voterAddress, caseId, salt)). Each wallet-case pair needs a unique 32-byte cryptographically generated salt, kept local and never reused; only its hash enters the receipt. Collision resistance relies on Keccak-256 [11].

A new salt would bypass a nullifier-only check, so hasVoted[caseId][voterAddress] must be set atomically on acceptance; usedNullifiers is secondary replay evidence. The reported tests and gas do not validate this strengthened rule.

Neither mapping prevents two enrolled wallets per person. Governance must enforce one active wallet, separated enrollment duties, an auditable identity registry and revocation log, and credential rotation that disables the old wallet first as shown in Fig. 3.

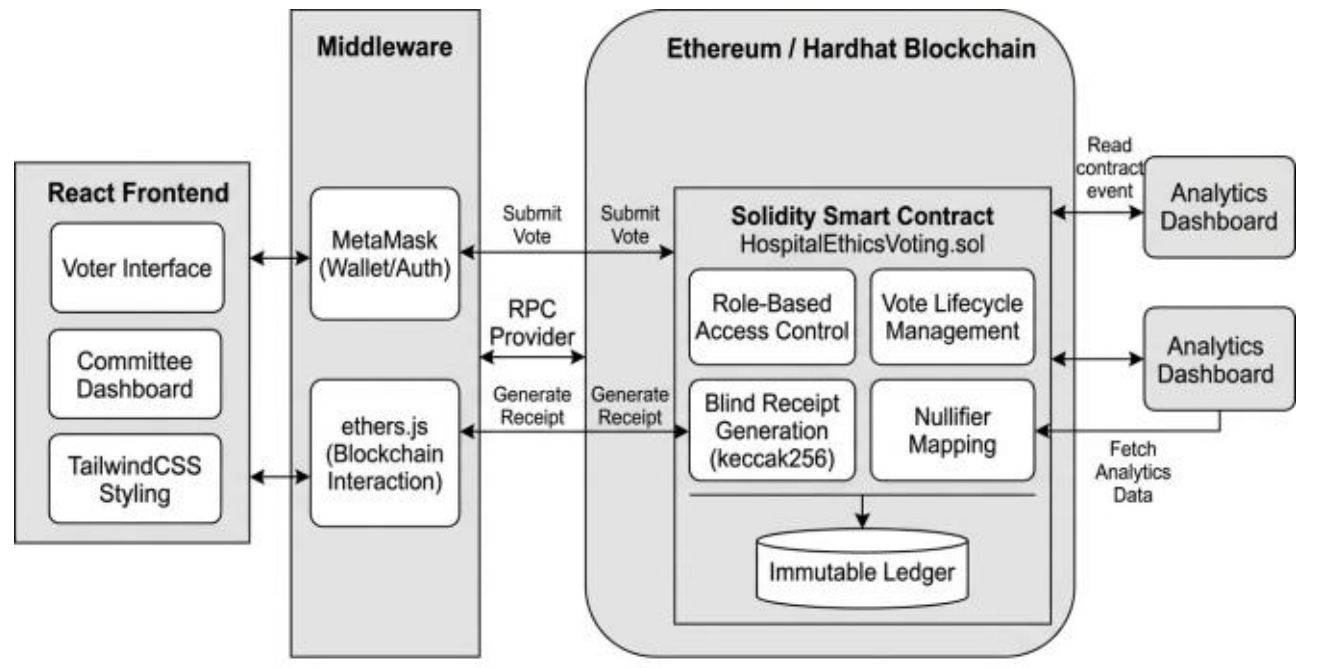

Fig. 3. Proposed system architecture.

## *D. Governance and access control*

Voters submit ballots, board members manage cases, and administrators enroll or revoke wallets. Because one administrator is a high-value failure point, production needs multisignature or two-person role changes, delayed emergency revocation, and independent registration review.

A malicious frontend can substitute call parameters. Human-readable wallet intent, reproducible builds, pinned contract addresses, integrity checks, an out-of-band case identifier, and an independent verifier reduce risk. Receipt recomputation detects post-confirmation mismatch but cannot prevent signing a substituted ballot.

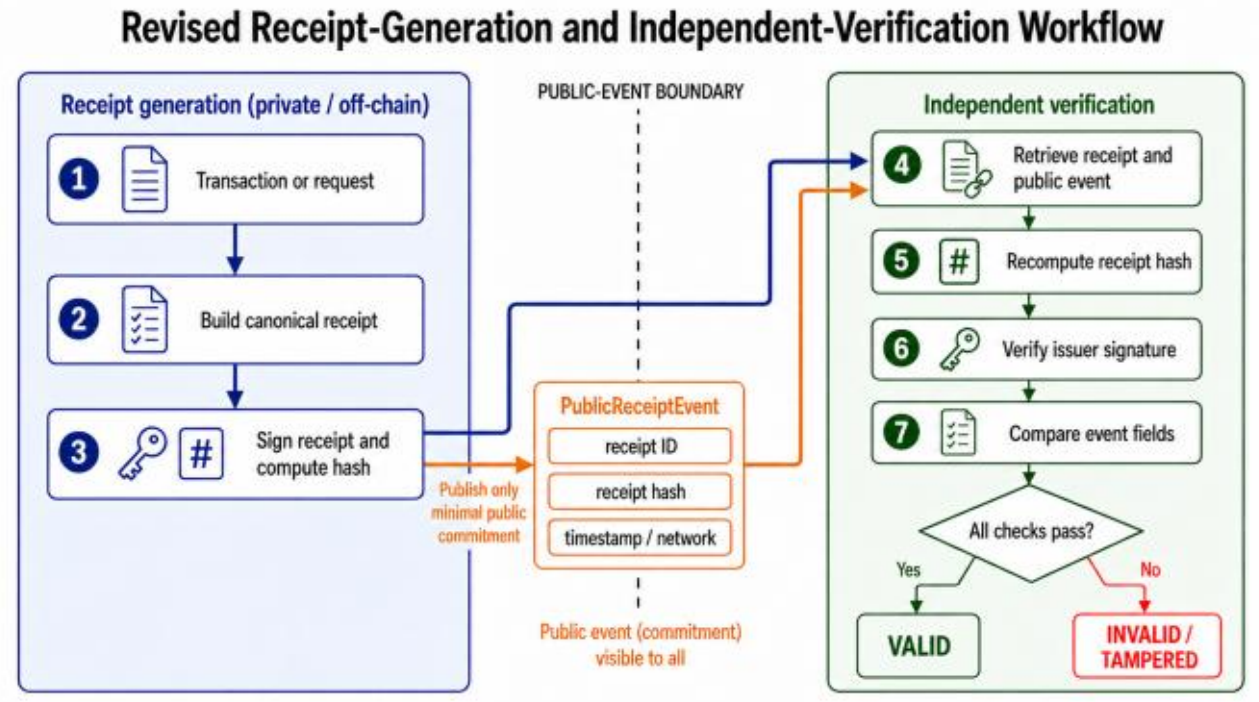

Fig. 4. Revised architecture with the off-chain identity registry, administrative trust boundary, and independent verifier made explicit.

## IV. THREAT MODEL AND SECURITY ANALYSIS

### A. Assets, actors, and assumptions

Assets are eligibility, ballots, case state, receipts, roles, credentials, and off-chain identity mappings. Adversaries are observers, enrolled voters, compromised clients, and compromised or colluding administrators. The ledger is assumed to order transactions and preserve confirmed history; voters protect keys, and case data are de-identified.

These assumptions give no secrecy: observers read addresses and votes; administrators may link identities; compromised clients may substitute ballots; and malicious administrators may register unauthorized or multiple wallets.

### B. Controls, evidence, and residual risk

TABLE II. SECURITY CLAIMS, AVAILABLE EVIDENCE, AND RESIDUAL RISK

| Threat/property | Control/evidence | Residual risk/status |
|---|---|---|
| Same-wallet duplicate | Nullifier test; address-case check | Partial evidence |
| Multiple wallets | Enrollment registry | Contract cannot prevent |
| Unauthorized caller | Role modifier; tests | Admin compromise |
| Invalid case state | Lifecycle checks claimed | Negative tests absent |
| Reentrancy | Guard claimed; class in [12] | Unaudited |
| Ballot substitution | Wallet + receipt check | Malicious client |
| Receipt tampering | Canonical recomputation | Needs all fields |
| Front-running | No control | Unmitigated |
| Gas/DoS | No stress test | Unmeasured |
| Ballot secrecy | Vote/address public | Not provided |
| Admin misuse | Role separation proposed | Single-point risk |

Table II avoids claiming threats are fully mitigated: functional tests are neither proofs, adversarial experiments, audits, nor evidence against administrator or frontend misuse.

### C. Required adversarial test matrix

The reported Hardhat suite omits the negative and adversarial cases in Table III. They must run against released code before deployment; none is reported as passing here.

TABLE III. REQUIRED NEGATIVE AND ADVERSARIAL TESTS

| Test | Expected invariant | Current evidence |
|---|---|---|
| Role bypass | Unauthorized call reverts | Not reported |
| Nullifier replay | Second call reverts atomically | Claim only |
| New salt/same address | Address-case check blocks | Not reported |
| Invalid case | No state/event change | Not reported |
| Front-run duplicate | At most one ballot | Not reported |
| Gas/large state | No state corruption | Not reported |
| Second wallet | Policy rejects or logs | Not implemented |
| Receipt tampering | Recomputation fails | Not reported |
| Credential rotation | Old wallet revoked first | Not reported |

## V. EVALUATION

### A. Evidence boundary and functional tests

Evidence comprises contract tests, local Hardhat gas units, and a synthetic workflow. Missing items include tool versions, compiler settings, hardware, repeated timing, bytecode hash, concurrency, state growth, and independent audit. Reproduction requires source, lockfile, deployment configuration, and logs.

TABLE IV. REPORTED FUNCTIONAL TEST COVERAGE

| Test suite | Tests passing |
|---|---|
| Hospital Ethics Voting | 12 / 12 |
| Receipt Verification | 10 / 10 |
| Total | 22 / 22 |

The 22 passing tests cover reported role, nullifier, receipt, and lifecycle paths; they do not establish secrecy, administrator or multiple-wallet resistance, ordering safety, or denial-of-service tolerance.

### B. Gas units and parametric monetary cost

TABLE V. REPORTED LOCAL HARDHAT GAS UNITS

| Action | Gas units |
|---|---:|
| Contract deployment | 1,821,618 |
| Verify voter | 47,515 |
| Create ethics case | 187,988 |
| Cast vote | 284,137 |
| Resolve case | 38,974 |

For gas G, price p (gwei), and ETH price P (USD), costUSD = $G \times p \times 10^{-9} \times P$. Table VI fixes P = USD 3,000 illustratively; it does not report current market prices.

TABLE VI. ILLUSTRATIVE TRANSACTION COST UNDER DEFINED ASSUMPTIONS (ETH = USD 3,000)

| Action | Gas | 1 gwei | 10 gwei | 30 gwei |
|---|---:|---:|---:|---:|
| Deployment | 1,821,618 | $5.46 | $54.65 | $163.95 |
| Verify voter | 47,515 | $0.14 | $1.43 | $4.28 |
| Create case | 187,988 | $0.56 | $5.64 | $16.92 |
| Cast vote | 284,137 | $0.85 | $8.52 | $25.57 |
| Resolve case | 38,974 | $0.12 | $1.17 | $3.51 |

Voting is the costliest recurring action. Local gas does not establish latency, throughput, finality, or affordability. Retained mappings imply at least O(V + C) growth, but bytes, execution time, concurrency, and storage scaling were not measured.

*C. Synthetic workflow simulation*

Twelve synthetic traces assume 70% receipt verification and 5% transaction failure; these are inputs, not observations. Simulated usability and trust scores cannot support deployment claims. A real, preregistered committee-member study should measure task success, errors, validated ratings, uncertainty, and qualitative findings.

*D. Scalability and state growth*

No test varies voters, cases, simultaneous submissions, or congestion. A complete benchmark must scale V and C; report repeated latency, throughput, verification time, failure rate, storage bytes, RPC load, and gas; and disclose gas limit, automining, confirmations, compiler, node, hardware, and network. Local Hardhat results cannot be generalized without these controls.

*E. Comparison with representative baselines*

Table VII compares an RBAC database, this prototype, and privacy-preserving designs [5]–[7], with Helios as open-audit context [9]. Only this prototype's gas is measured here; cross-paper timing and gas are incomparable across platforms, hardware, ballots, and guarantees.

TABLE VII. EVIDENCE-BOUNDED COMPARISON WITH CENTRALIZED AND PRIVACY-PRESERVING APPROACHES

| Metric | Centralized RBAC database | Proposed prototype | Privacy-preserving blockchain [5]–[7] |
|---|---|---|---|
| Eligibility | Server registry | Admin-enrolled wallet + role | Credential, blind signature, or proof |
| Ballot confidentiality | Depends on operator controls | Not provided; vote/address public | Encrypted ballot and/or anonymous credential |
| Auditability | Operator logs/audit | Public events + receipt | Public cryptographic proofs |
| Duplicate-vote control | Unique constraint | Address-case + nullifier | One-time credential or proof |
| Per-ballot cost | Not measured | 284,137 local gas | No same-platform measure |
| Verification work | Query/log review | Event + ABI/Keccak-256 | Proof/signature or decryption |
| Storage growth | O(V) off-chain | At least O(V) on-chain | O(V) + cryptographic artifacts |
| Implementation complexity | Low to moderate | Moderate | High |
| Same-platform benchmark | Not conducted | Reference only | Not conducted |

The prototype reduces cryptographic complexity by exposing ballot and identity linkage; Table VII is not a performance claim. A valid benchmark must hold compiler, network, hardware, ballot size, case count, and confirmations constant while reporting computation, storage, gas, verification time, and guarantees.

## VI. DISCUSSION AND LIMITATIONS

The design may suit transparent, low-frequency polls where linked vote visibility is acceptable; it is unsuitable for confidential ethics voting. Events, enrollment records, browser storage, and transaction metadata provide linkage paths.

A receipt proves only that specified public fields occur on a selected chain. It cannot prove correct enrollment, one wallet per person, honest administration, secrecy, or that the frontend showed the signed choice. Detection after confirmation does not reverse substitution.

Material limits are absent adversarial coverage, local gas without performance evidence, assumption-driven usability data, and no controlled baseline implementation. Hospital readiness is not established.

## VII. FUTURE WORK

Priorities are reproducible code, execution of Table III, atomic address-case uniqueness, multisignature enrollment, auditable credential rotation, and independent contract/frontend review. Privacy needs encrypted commitments, threshold tallying or zero knowledge, plus front-running controls.

Future benchmarks should scale voters and cases while measuring latency, finality, throughput, RPC load, state growth, and cost under documented conditions and identical baselines. Real committee-member studies should test wallet-prompt comprehension, substitution detection, verification success, time, errors, workload, and calibrated trust.

## VIII. CONCLUSION

The prototype combines role-controlled workflow, duplicate-vote checks, and inclusion receipts for small ethics committees. It offers pseudonymous public auditability, not secret voting, and evidence is limited to reported functions and local gas. Confidential deployment requires governed identity, encrypted ballots, adversarial tests, reproducible benchmarks, independent audit, and real users.

### Acknowledgment of Generative AI Use

The authors did not use generative AI to draft or generate the scientific content of this article. A generative AI tool was used solely for grammar correction and English-language proofreading.